\documentclass[amsmath,aps,prd,reprint,showpacs,groupedaddress]{revtex4-1}

\usepackage{graphicx}
\usepackage{tensor}

\newcommand{\intd}[2][]{\: \mathrm{d}^{#1} #2 \:}
\newcommand{\pd}[2][]{\operatorname{\partial}_{#2}^{#1}}

\newcommand{\defeq}{\mathrel{\vcenter{\baselineskip0.5ex \lineskiplimit0pt
                     \hbox{\scriptsize.}\hbox{\scriptsize.}}}%
                     =}
\newcommand{\eqdef}{=\mathrel{\vcenter{\baselineskip0.5ex \lineskiplimit0pt
                     \hbox{\scriptsize.}\hbox{\scriptsize.}}}%
                      }
\DeclareMathOperator{\diag}{diag}

\begin{document}

\title{Inverse approach to solutions of the Dirac equation for space-time 
dependent fields}

\author{Johannes Oertel}
\author{Ralf Sch\"utzhold}
\email{ralf.schuetzhold@uni-due.de}
\affiliation{Fakult\"at f\"ur Physik, Universit\"at Duisburg-Essen, 
Lotharstrasse 1, 47057 Duisburg, Germany}

\date{\today}

\begin{abstract}
Exact solutions of the Dirac equation in external electromagnetic background 
fields are very helpful for understanding non-perturbative phenomena in 
quantum electrodynamics (QED). 
However, for the limited set of known solutions, the field often depends on 
one coordinate only, which could be the time $t$, a spatial coordinate 
such as $x$ or $r$, or a light-cone coordinate such as $ct-x$.
By swapping the roles of known and unknown quantities in the Dirac equation, 
we are able to generate families of solutions of the Dirac equation in the 
presence of genuinely space-time dependent electromagnetic fields in $1+1$ 
and $2+1$ dimensions.
\end{abstract}


\maketitle

\section{Introduction}

Quantum electrodynamics (QED) as the theory of charged particles interacting 
with electromagnetic fields is well understood in the context of standard 
perturbation theory and can describe several intriguing phenomena of nature. 
However, QED contains other fascinating effects that cannot be explained 
using perturbative methods. 
Such non-perturbative effects can arise when the electromagnetic field is so 
strong that it cannot be treated as a perturbation.
In order to understand these phenomena, it is often useful to study the 
behaviour of exact solutions in those external background fields. 

Unfortunately, although the Dirac equation was first formulated more than 
eighty years ago \cite{Dirac1928,*Dirac1930}, the set of known exact solutions 
is still quite limited (see, e.g., \cite{Bagrov2014} for a review).
Apart from the Coulomb field $\propto1/r^2$
\cite{Gordon1928,Darwin1928}, exact solutions are known for a constant
electric field and a Sauter profile in space $\propto1/\cosh^2(kx)$ or
time $\propto1/\cosh^2(\omega t)$, for example.
The latter are relevant for the non-perturbative Sauter-Schwinger effect 
\cite{Sauter1931,*Sauter1932,Heisenberg1936,Schwinger1951} corresponding to 
electron-positron pair creation from vacuum via tunnelling. 
In contrast to electron-positron pair creation in the perturbative
(multi-photon) regime which has been observed at SLAC \cite{Burke1997}, this
non-perturbative prediction of quantum field theory has not been conclusively
experimentally verified yet.
However, there are several experimental initiatives which might be able to
eventually reach the ultra-strong field regime necessary for observing this
striking effect \cite{Note1}.

Furthermore, exact solutions are known for a constant magnetic field
(relativistic Landau levels, see \cite{Rabi1928,Canuto1969}) and plane
waves, where the fields depend on one of the light-cone coordinates
such as $ct-x$ (Volkov solutions, see
e.g. \cite{Volkov1935,Nikishov1964,Tomaras2000,*Tomaras2001,Hebenstreit2011b}).
These (transverse) fields do not induce pair creation from vacuum.

Nevertheless, in all these cases, the fields depend on one coordinate only 
(such as $r$, $x$, $t$, or $ct-x$). 
As a result of this high degree of symmetry, the set of partial
differential equations can be reduced to an ordinary differential
equation, which greatly simplifies the analysis.
An analogous limitation applies to our theoretical understanding of the 
Sauter-Schwinger effect. 
Even though there are many results for fields which depend on one coordinate
only, we are just beginning to understand the impact of the interplay between
spatial and temporal dependencies, see, e.g., \cite{Linder2015, Schneider2015,
  Hebenstreit2010, Hebenstreit2011a, Hebenstreit2011c, Ruf2009}.

In the following, we develop a method which allows us to obtain solutions 
of the Dirac equation for genuinely space-time dependent fields. 
To this end, we pursue a different approach by assuming that we already know 
a solution to the Dirac equation. 
We then calculate the vector potential $A_\mu$ corresponding to the given 
solution from the Dirac equation. 
This is feasible as the Dirac equation does not contain any derivatives of 
the vector potential. 
More generally speaking, we write down a solution to a partial differential 
equation and then try to find a physical problem associated with the solution 
-- a concept also well known in the field of fluid dynamics, see, for example, 
\cite{Nemenyi1951}.

\section{Light cone coordinates}
\label{sec:lightcone}

Let us start with the most simple and yet non-trivial case -- 
the Dirac equation in 1+1 dimensions. 
For the following derivation, it is convenient to transform to 
light cone coordinates $x_\pm$ defined as ($\hbar=c=1$)
\begin{equation}
  x_+ = \frac{t + x}{\sqrt{2}}, \quad x_- = \frac{t - x}{\sqrt{2}}.
\end{equation}
Since perhaps not all readers will be familiar with the form of the 
subsequent expressions in light cone coordinates, let us insert a brief 
reminder. 
The Jacobian matrix of the coordinate transformation between Cartesian 
and light cone coordinates 
\begin{equation}
J\indices{^\mu_\nu} 
= 
\frac{\partial \left( x_+, x_- \right)}{\partial \left( t, x \right)} 
= 
\frac{1}{\sqrt{2}} \begin{pmatrix}1 & 1 \\ 1 & -1 \end{pmatrix},
\end{equation}
yields the transformation laws for tensors such as the partial derivatives 
\begin{equation}
{\pd{\mu}}' 
= 
(J^{-1})\indices{^\nu_\mu} \pd{\nu}
= 
\frac{1}{\sqrt{2}} \begin{pmatrix} \pd{t}+\pd{x} \\ \pd{t}-\pd{x} \end{pmatrix}
= 
\begin{pmatrix} \pd{x_+} \\ \pd{x_-} \end{pmatrix}
\eqdef \begin{pmatrix} \pd{+} \\ \pd{-} \end{pmatrix}.
\end{equation}
In 1+1 dimensions, the electromagnetic field strength tensor contains only 
one independent component, the electric field $E(t,x)$ 
\begin{equation}
F_{\mu \nu} 
= 
\pd{\mu} A_\nu - \pd{\nu} A_\mu 
= 
\begin{pmatrix} 0 & E \\ -E & 0 \end{pmatrix},
\end{equation}
which thus reads in light cone coordinates
\begin{equation}
F'_{\mu \nu} 
= 
(J^{-1})\indices{^\lambda_\mu} (J^{-1})\indices{^\rho_\nu} F_{\lambda \rho} 
= 
\begin{pmatrix}0 & -E \\ E & 0 \end{pmatrix}.
\label{eq:fieldtensor1p1}
\end{equation}
Transforming the Cartesian Minkowski metric tensor 
$\eta_{\mu \nu} = \diag(+1, -1)$ to light cone coordinates as well gives
\begin{equation}
  g_{\mu \nu} = \begin{pmatrix}0 & 1 \\ 1 & 0 \end{pmatrix}.
\end{equation}
A possible choice of light cone gamma matrices satisfying the Clifford
algebra's anti-commutation relation
\begin{equation}
  \left\{ \gamma^\mu, \gamma^\nu \right\} = 2 g^{\mu \nu},
\end{equation}
therefore is
\begin{equation}
\gamma^+ 
= 
\begin{pmatrix}0 & \sqrt{2} \\ 0 & 0\end{pmatrix}, \quad \gamma^- 
= 
\begin{pmatrix}0 & 0 \\ \sqrt{2} & 0\end{pmatrix}.
\label{eq:lightconegamma}
\end{equation}
Note that in 1+1 and 2+1 dimensions, the Clifford algebra can be 
satisfied with $2\times2$-matrices. 

\section{Inverse approach}
\label{sec:inverseapproach}

The Dirac equation, minimally coupled to the electromagnetic potential $A_\mu$
via the charge $q$
\begin{equation}
  \left( i \gamma^\mu \left[ \pd{\mu} + i q A_\mu \right] - m \right) \psi = 0,
\end{equation}
assumes the following simple form in terms of the light cone gamma matrices 
\eqref{eq:lightconegamma}
\begin{equation}
\begin{pmatrix} -m & i \sqrt{2} \left[ \pd{+} + i q A_+ \right] \\ 
i \sqrt{2} \left[ \pd{-} + i q A_- \right] & -m \end{pmatrix} 
\begin{pmatrix} \psi_1 \\ \psi_2 \end{pmatrix} = 0.
  \label{eq:lightconediraceq}
\end{equation}
Traditionally, the Dirac equation is treated as a partial differential
equation. 
A solution $\psi$ for a specific potential $A_\mu$ is typically
calculated by reducing the Dirac equation to an ordinary differential
equation. 
In our approach, we assume that we know a specific spinor 
$\psi = (\psi_1, \psi_2)^\mathsf{T}$ which is a solution to the Dirac
equation and calculate the corresponding potential. 
Thus, we solve \eqref{eq:lightconediraceq} for the components of $A_\mu$
\begin{equation}
  \begin{aligned}
    q A_+ 
&= 
i \frac{\pd{+} \psi_2}{\psi_2} - \frac{m}{\sqrt{2}} \frac{\psi_1}{\psi_2}, 
\\
    q A_- 
&= 
i \frac{\pd{-} \psi_1}{\psi_1} - \frac{m}{\sqrt{2}} \frac{\psi_2}{\psi_1}.
  \end{aligned}
\end{equation}
For arbitrary $\psi$, these expressions are not necessarily real. 
Therefore, we require the imaginary parts of $q A_+$ and $q A_-$ to vanish, 
giving two conditions which we use to eliminate two real degrees of freedom 
of the spinor $\psi$. 
Using the polar representation for the spinor components 
$\psi_k = r_k e^{i \varphi_k}$, these conditions can be written as
\begin{equation}
  \begin{aligned}
r_2 \pd{+} r_2 - \frac{m}{\sqrt{2}}\, r_1 r_2 
\sin \left( \varphi_1 - \varphi_2 \right) &= 0, \\
r_1 \pd{-} r_1 + \frac{m}{\sqrt{2}}\, r_1 r_2 
\sin \left( \varphi_1 - \varphi_2 \right) &= 0.
  \end{aligned}
\end{equation}
Adding the two equations gives 
\begin{subequations}
  \begin{align}
    \pd{-} r_1^2 &= - \pd{+} r_2^2 \label{eq:radial},\\
    \pd{-} r_1 &= \frac{m}{\sqrt{2}}\, r_2 
\sin \left( \varphi_2 - \varphi_1 \right) \label{eq:phase}.
  \end{align}
\end{subequations}
The first equation \eqref{eq:radial} can be solved for $r_2$ by integrating 
with respect to $x_+$
\begin{equation}
  r_2 = \sqrt{c(x_-) - \int \pd{-} r_1^2 \intd{x_+}},
  \label{eq:r2}
\end{equation}
where $c(x_-)$ is an integration constant that may still depend on $x_-$. 
The remaining equation \eqref{eq:phase} determines the phase difference 
$\varphi_2 - \varphi_1$ 
\begin{equation}
\label{eq:phi2}
\varphi_2 - \varphi_1 
= 
\arcsin \left( \frac{\sqrt{2}}{m} \frac{\pd{-} r_1}{r_2} \right),
\end{equation}
where we could also use other branches of the $\arcsin$-function such as
$\Delta\varphi=\varphi_2 - \varphi_1\to\pi-\Delta\varphi$, leading to different
solutions in general -- see the remark after Eq.~\eqref{eq:spinor}.
Using the abbreviation 
\begin{equation}
s = \sqrt{c - \int \pd{-} r^2 \intd{x_+} - \frac{2}{m^2} (\pd{-} r)^2},
\label{eq:s}
\end{equation}
and Eqs.~\eqref{eq:r2} and \eqref{eq:phi2}, we can calculate the form of the 
spinor $\psi$
\begin{equation}
  \psi 
= 
\begin{pmatrix} \psi_1 \\ \psi_2 \end{pmatrix}
= 
e^{i \varphi} 
\begin{pmatrix} r \\ \pm s + i \frac{\sqrt{2}}{m} \pd{-}r \end{pmatrix},
  \label{eq:spinor}
\end{equation}
where we have set $r = r_1$ and $\varphi = \varphi_1$. 
Note that we find two different solutions with $\pm s$ corresponding to the 
different branches of the $\arcsin$ or square-root functions in 
Eqs.~\eqref{eq:phi2} and \eqref{eq:s}, respectively. 

Local gauge invariance allows us to eliminate the phase $e^{i \varphi}$ by
applying a gauge transformation 
$\psi \mapsto \psi' = e^{-i \varphi} \psi$, which adds a term 
$\pd{\mu}\varphi$ to $q A_\mu$. 
The components of $A_\mu$ using the spinor given in \eqref{eq:spinor}
finally are
\begin{equation}
  \begin{aligned}
q A_+ 
&= 
\mp\frac{m}{\sqrt{2}}\frac{r}{s}\mp\frac{\sqrt{2}}{m}\frac{\pd{+}\pd{-}r}{s}, 
\\
q A_- 
&= 
\mp \frac{m}{\sqrt{2}} \frac{s}{r}.
\end{aligned}
\label{eq:qA}
\end{equation}
These are obviously real as long as $r$ and $s$ are real, too. 
The electric field corresponding to this potential according to
\eqref{eq:fieldtensor1p1} is
\begin{equation}
  E = \pd{-} A_+ - \pd{+} A_-.
  \label{eq:E}
\end{equation}
In summary, by choosing a real generating function $r(x_+, x_-)$ and a real  
supplementary boundary value function $c(x_-)$, we can generate arbitrary 
space-time dependent solutions $\psi(x_+, x_-)$ of the Dirac equation in the 
presence of an electromagnetic background $A_\mu$, which can also depend 
on space and time. 

Obviously, the associated electromagnetic field strength tensor $F^{\mu \nu}$ in
Eq.~\eqref{eq:fieldtensor1p1} automatically satisfies the homogeneous Maxwell
equations as it has been derived from a vector potential $A_\mu$.
If we demand that it also obeys the inhomogeneous Maxwell equations
($\mu_0=1$)
\begin{equation}
\pd{\nu} F^{\mu \nu} = j^\mu
\,,
\end{equation}
we have to specify the sources $j^\mu$ accordingly. 
In $1+1$ dimensions we find
\begin{equation}
\begin{aligned}
\rho &=&-&\pd{x} E &&= \frac{1}{\sqrt{2}} \left( \pd{-} E - \pd{+} E \right), 
\\
j &=& &\pd{t} E &&= \frac{1}{\sqrt{2}} \left( \pd{-} E + \pd{+} E \right),
\end{aligned}
\end{equation}
where $\rho$ is the charge density and $j$ is the current density.
For non-trivial field profiles $E(t,x)$, they will be non-zero in general.
However, this is no surprise because the only vacuum solution of the 
Maxwell equations in $1+1$ dimensions  is a constant electric field
$E=\rm const$.

\section{Solutions}
\label{sec:solutions}

In order to illustrate the approach presented in the previous section, let us
discuss some exemplary solutions that can be found using this method, starting
with the most simple ones.
The expressions for the spinor and the potential components are 
significantly simplified if $r$ is independent of $x_-$.

\subsection{Plane waves}\label{Plane waves}

Choosing $r$ and $s=\pm\sqrt{c}$ to be constant,
\begin{equation}
  \psi = \begin{pmatrix} r \\ s \end{pmatrix} = \mathrm{const},
\end{equation}
leads to a constant electromagnetic vector potential
\begin{equation}
  \begin{aligned}
    q A_+ &= -\frac{m}{\sqrt{2}} \frac{r}{s} = \mathrm{const}, \\
    q A_- &= -\frac{m}{\sqrt{2}} \frac{s}{r} = \mathrm{const}.
  \end{aligned}
\end{equation}
Thus, a gauge transformation $\psi\mapsto\psi' = e^{-i p_\mu x^\mu} \psi$ with
\begin{equation}
  p_\mu \defeq 
\begin{pmatrix} p_+ \\ p_- \end{pmatrix} 
= 
\frac{m}{\sqrt{2}} \begin{pmatrix} r/s \\ s/r \end{pmatrix}
\end{equation}
can be used to set the potential components to zero and reveals that
these solutions are plane wave solutions to the free Dirac equation of
either positive or negative energy. 
Transforming the light-cone momenta $p_\pm$ back to the usual Cartesian 
representation $p_0=(p_++p_-)/\sqrt{2}$, we find that the energy is given 
by $p_0=m(r/s+s/r)/2$.
Thus a solution where both $r$ and $s$ are positive (or both negative) 
corresponds to a positive energy whereas different signs of $r$ and $s$
yield a negative energy. 

\subsection{Single pulses}

In this subsection, we find solutions for arbitrary light cone fields
$E(x_+)$ and $E(x_-)$, i.e., pulses moving along the light lines. 
Such solutions were found before using traditional methods as well
\cite{Tomaras2000,*Tomaras2001,Hebenstreit2011b}.

\subsubsection{$x_+$-dependent pulse}

Let us assume that the function $r$ depends on $x_+$ only while 
$s = \pm\sqrt{c} = \mathrm{const}$
\begin{equation}
  \psi = \begin{pmatrix}r(x_+) \\ s \end{pmatrix}.
\end{equation}
In this case, neither the spinor nor the vector potential depends on 
$x_-$ which simplifies the expression for the electric field 
\begin{equation}
\label{electric-field}
q E 
=
q \underbrace{\pd{-} A_+}_{= 0} - q \pd{+} A_- 
= 
\frac{m}{\sqrt{2}} s \pd{+} \frac{1}{r(x_+)}.
\end{equation}
This is a first-order ordinary differential equation for $r(x_+)$
which can be integrated easily 
\begin{equation}
r(x_+) 
= 
r_\mathrm{in}
\left[
{1 + \frac{\sqrt{2}}{m} \frac{r_\mathrm{in}}{s} q 
\int_{-\infty}^{x_+} E(\tilde{x}_+) \intd{\tilde{x}_+}}
\right]^{-1}
,
\label{eq:singlepr}
\end{equation}
with $r_\mathrm{in} = r(x_+ \to -\infty)$.
Comparison with Sec.~\ref{Plane waves} reveals that the pre-factor 
$\sqrt{2}r_\mathrm{in}/(ms)$ in front of the above $x_+$-integral over $qE$ 
is just the inverse initial momentum $1/p_-^\mathrm{in}$.  
As already discussed in \cite{Tomaras2000}, the term in the square bracket
in Eq.~\eqref{eq:singlepr} vanishes and thus $r$ diverges when this 
$x_+$-integral over $qE$ becomes large enough to compensate $p_-^\mathrm{in}$.  
Note that the light cone dispersion relation $p_+p_-=m^2/2$ shows that 
$p_+$ must diverge when $p_-$ vanishes and vice versa. 

Now, let us recall that the phenomenon of electron-positron pair creation 
(such as in the Sauter-Schwinger effect) can be described by the situation 
where an initial solution with positive energy transforms into a final 
solution which contains contributions with negative energies (or vice versa).
Assuming that $r$ becomes constant initially and finally, we find that pair 
creation can only occur if $r(x_+)$ changes its sign somewhere, i.e., 
if $r(x_+)$ vanishes or diverges at some point. 
If $r(x_+)$ crosses zero, the electric field~\eqref{electric-field} 
diverges -- whereas a diverging $r(x_+)$ precisely corresponds to the 
case discussed above, see also \cite{Tomaras2000}.
Thus, we find that we cannot describe particle creation in this case 
without introducing some singularity (see the Appendix). 

\subsubsection{$x_-$-dependent pulse}

In a similar way, we can derive solutions for electric fields only depending 
on $x_-$ by setting $r=\mathrm{const}$ and letting $s(x_-)=\pm\sqrt{c(x_-)}$ 
depend on $x_-$
\begin{equation}
  \psi = \begin{pmatrix}r \\ s(x_-) \end{pmatrix}.
\end{equation}
Thus, the electric field can be calculated as follows
\begin{equation}
q E 
= 
q \pd{-} A_+ - q \underbrace{\pd{+} A_-}_{= 0} 
= 
- \frac{m}{\sqrt{2}} r \pd{-} \frac{1}{s(x_-)},
\end{equation}
which is again a first-order ordinary differential equation for
$s(x_-)$. 
The solution is given by
\begin{equation}
s(x_-) 
= 
s_\mathrm{in}
\left[
1 - \frac{\sqrt{2}}{m} \frac{s_\mathrm{in}}{r} 
q \int_{-\infty}^{x_-} E(\tilde{x}_-) \intd{\tilde{x}_-}
\right]^{-1}
,
\end{equation}
with $s_\mathrm{in} = s(x_- \to -\infty)$.
In complete analogy, the same arguments as for an $x_+$-dependent pulse apply 
in this case.  

\subsection{Two pulses}

As a non-trivial extension of these two cases, we can combine the previous 
two solutions into a single spinor
\begin{equation}
  \psi = \begin{pmatrix}r(x_+) \\ s(x_-) \end{pmatrix},
\end{equation}
where the two components are given by 
\begin{equation}
  \begin{aligned}
    r(x_+) 
&= 
r_\mathrm{in}
\left[
1 + \frac{\sqrt{2}}{m} \frac{r_\mathrm{in}}{s_\mathrm{in}} 
q \int_{-\infty}^{x_+} E_+(\tilde{x}_+) \intd{\tilde{x}_+}
\right]^{-1}
, 
\\
    s(x_-) 
&= 
s_\mathrm{in}
\left[
1 - \frac{\sqrt{2}}{m} \frac{s_\mathrm{in}}{r_\mathrm{in}} 
q \int_{-\infty}^{x_-} E_-(\tilde{x}_-) \intd{\tilde{x}_-}
\right]^{-1}
.
  \end{aligned}
\end{equation}
We may calculate the electric field using \eqref{eq:qA} and \eqref{eq:E}
\begin{equation}
E(x_+,x_-) 
= 
\frac{r(x_+)}{r_\mathrm{in}} E_-(x_-) + \frac{s(x_-)}{s_\mathrm{in}} E_+(x_+).
\end{equation}
Initially, we have $E(x_+,x_-)=E_-(x_-)+E_+(x_+)$ which corresponds to two 
independent pulses approaching each other from different directions. 
When these two pulses meet, however, this is no longer true -- which shows 
that the mapping from $\psi$ (i.e., $r$ and $s$) to $A_\mu$ is not linear. 
For late times, these pulses propagate again independently, but with 
modified amplitudes in general. 

\subsection{Emerging pulses}

\begin{figure}
  \includegraphics{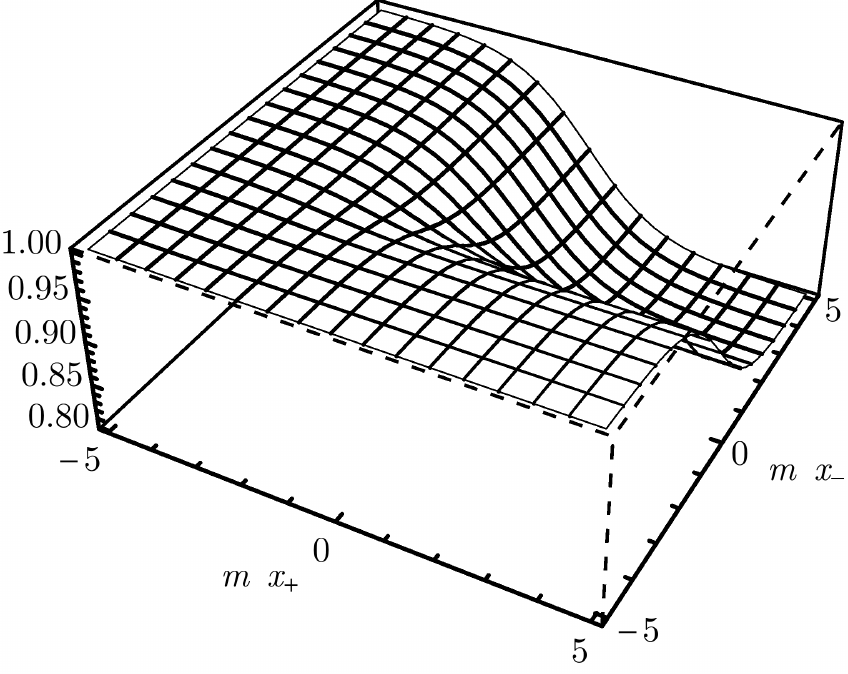}
  \caption{\label{fig:emergingr}Plot of $r(x_+, x_-)$ as given in
    \eqref{eq:emergingr} with $r_\mathrm{in} = 1$, $\xi = 0.2$, and
    $\gamma = 1.2/m$.}
\end{figure}

Another solution where the corresponding electric field consists of
two pulses can be generated by setting
\begin{equation}
  r(x_+, x_-) = r_\mathrm{in} + \frac{\xi}{1 + e^{-\gamma x_+} + e^{-\gamma x_-}}.
  \label{eq:emergingr}
\end{equation}
For non-vanishing $\xi$ and $\gamma > 0$, the chosen $r(x_+, x_-)$ will be
constant almost everywhere except in the vicinity of the forward light
cone (see figure \ref{fig:emergingr}).

\begin{figure}
  \includegraphics{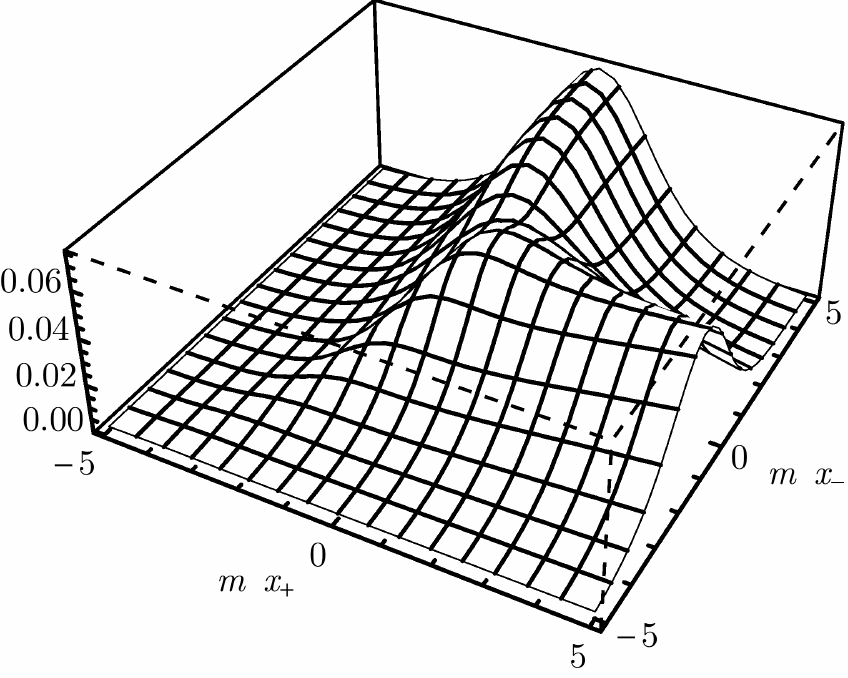}
  \caption{\label{fig:emerginge}Plot of the electric field $qE$
    corresponding to the solution generated by $r(x_+, x_-)$ given in
    \eqref{eq:emergingr} with $r_\mathrm{in} = 1$, $\xi = 0.2$, and
    $\gamma = 1.2/m$.}
\end{figure}

In this case, the expression for $s$ according to \eqref{eq:s} is not
as simple as before because $r$ is not independent of $x_-$. 
Nevertheless, $s$ can be calculated analytically, although the
resulting expressions for $s$ and the electric field $q E$ are quite
lengthy. 
Thus, we will only give a plot of the resulting electric field which shows 
the two pulses emerging from the origin and moving along the forward light 
lines (see figure \ref{fig:emerginge}).

\section{Extension to 2+1 dimensions}
\label{sec:2d}

The approach presented here can be extended to $2+1$ dimensional space-times 
as well. 
We use the Cartesian coordinate $y$ in addition to the light cone coordinates 
$x_+$ and $x_-$. 
Thus, the metric tensor becomes
\begin{equation}
g\indices{^{\mu \nu}} 
= 
\begin{pmatrix} 0 & 1 & 0 \\ 1 & 0 & 0 \\ 0 & 0 & -1 \end{pmatrix}.
\end{equation}
In order to complete our set of gamma matrices from \eqref{eq:lightconegamma}, 
we choose the third $\gamma$-matrix according to 
\begin{equation}
  \gamma^2 = i \sigma_z = \begin{pmatrix} i & 0 \\ 0 & -i \end{pmatrix}. 
\end{equation}
Thus the Dirac equation in $2+1$ dimensions is given by
\begin{equation}
  \begin{pmatrix} 
-m - \left[ \pd{y} + i q A_y \right] 
& i \sqrt{2} \left[ \pd{+} + i q A_+ \right] 
\\ i \sqrt{2} \left[ \pd{-} + i q A_- \right] 
& -m + \left[ \pd{y} + i q A_y \right] \end{pmatrix} 
\begin{pmatrix} \psi_1 \\ \psi_2 \end{pmatrix} = 0.
\end{equation}
In complete analogy to section \ref{sec:inverseapproach}, we solve the
Dirac equation for $q A_+$ and $q A_-$ and reduce the spinor's number
of degrees of freedom by requiring the imaginary parts of the
electromagnetic potential's components to vanish. 
After some calculation, we are able to write the spinor and the 
electromagnetic potential in terms of three real functions 
$r_1(x_+, x_-, y)$, $r_2(x_+, x_-, y)$ and $c(x_+,x_-)$. 
Explicitly, a spinor of the form
\begin{equation}
\psi 
= 
\begin{pmatrix} 
r_1 
\\
s - i u 
\end{pmatrix},
\end{equation}
with
\begin{equation}
   s = \pm \sqrt{r_2^2 - u^2}
\end{equation}
and
\begin{equation}
u 
= 
\frac{1}{\sqrt{2} r_1} \left[
c(x_+,x_-) + \int \left(\pd{-} r_1^2 + \pd{+} r_2^2 \right) \intd{y} 
\right]
\end{equation}
is a solution of the Dirac equation with the potential components
\begin{equation}
\begin{aligned}
q A_+ 
&= 
- \frac{m}{\sqrt{2}} \frac{r_1}{s} 
- \frac{1}{\sqrt{2}} \frac{\pd{y} r_1}{s} 
+ \frac{\pd{+} u}{s}, 
\\ 
q A_- 
&= 
- \frac{m}{\sqrt{2}} \frac{r_2^2}{r_1 s} 
+ \frac{1}{\sqrt{2}} \frac{r_2 \pd{y} r_2}{r_1 s} 
- \frac{u \pd{-} r_1}{r_1 s}, 
\\
q A_y 
&= 
- m \frac{u}{s} - \frac{u \pd{y} r_1}{r_1 s} 
+ \frac{1}{\sqrt{2}} \frac{\pd{+} r_2^2}{r_1 s}.
\end{aligned}
\end{equation}
In contrast to 1+1 dimensions, the electromagnetic field strength tensor
contains three independent components, for example the two electric fields 
$E_{x,y}$ in $x$ and $y$ direction plus the perpendicular magnetic field $B_z$. 
These components of the electromagnetic field can be calculated as follows
\begin{equation}
\begin{aligned}
E_x &= \pd{-} A_+ - \pd{+} A_-, 
\\
E_y &= \frac{1}{\sqrt{2}} 
\left( \pd{-} A_y - \pd{y} A_- + \pd{+} A_y - \pd{y} A_+ \right), 
\\
B_z &= \frac{1}{\sqrt{2}} 
\left( \pd{-} A_y - \pd{y} A_- - \pd{+} A_y + \pd{y} A_+ \right).
  \end{aligned}
\end{equation}
We see that these expressions simplify significantly if $r_1$ and $r_2$ are
independent of $y$. 
In that case, the electromagnetic field does only depend on the light cone 
coordinates as before and similar solutions as in the $1+1$ dimensional case 
can be found, e.g. one and two wavefronts. 
In fact, the solutions given in section \ref{sec:solutions} are solutions 
to the $2+1$ dimensional Dirac equation as well but can be extended to also 
include a transverse electric and magnetic field component.

To verify that our method reproduces known solutions, we insert the
lowest Landau level solution 
\begin{equation}
\psi 
= 
\mathcal{N} \exp 
\left( -\frac{1}{2} q B \left[ x - \frac{k_y}{q B} \right]^2 \right) 
\begin{pmatrix} 1 \\ 1 \end{pmatrix}
\end{equation}
into our formalism, i.e. we set
\begin{equation}
\begin{aligned}
r_1(x_+, x_-) 
&= 
\mathcal{N} \exp\left(
-\frac{1}{2}qB\left[\frac{x_+-x_-}{\sqrt{2}}-\frac{k_y}{q B}\right]^2
\right), 
\\
r_2(x_+, x_-) 
&= 
r_1(x_+, x_-), \quad c = 0,
\end{aligned}
\end{equation}
where $\mathcal{N}$ is a normalization constant. 
Calculating the potential components gives
\begin{equation}
q A_+ = q A_- = -\frac{m}{\sqrt{2}}, 
\ q A_y = - q B \frac{x_+ - x_-}{\sqrt{2}} + k_y,
\end{equation}
so that the electromagnetic field is
\begin{equation}
  E_x = E_y = 0, \ B_z = B,
\end{equation}
which is the expected result.

\section{Conclusions \& Outlook}

We have developed an inverse approach for generating families of exact 
solutions of the Dirac equation in the presence of space-time dependent 
electromagnetic fields in 1+1 and 2+1 dimensions. 
Somewhat similar to optimal control theory, we start with a suitable ansatz 
for the spinor $\psi$ and then derive the appropriate background field $A_\mu$ 
which supports this solution.
In 1+1 dimensions, we may choose a real generating function $r(x_+, x_-)$ 
and a suitable real supplementary boundary value function $c(x_-)$ such that 
the radicand in Eq.~\eqref{eq:r2} stays positive. 
In 2+1 dimensions, we may choose two real generating functions 
$r_1(x_+, x_-, y)$ and $r_2(x_+, x_-, y)$ as well as one real boundary value 
function $c(x_+,x_-)$. 

The solutions generated in this way may depend on space and time in a 
complicated manner -- a situation which is quite difficult to treat with 
traditional methods.
As one possible application, our method could be used to solve steering 
problems such as: given an initial wave-packet $\psi_{\rm in}$, which 
electromagnetic field $A_\mu$ induces an evolution to a prescribed final 
wave-packet $\psi_{\rm out}$?
As another application, these exact solutions could be used as touchstones for
already existing exact or approximate non-perturbative derivation techniques
(e.g. the worldline instanton method \cite{Dunne2005}) or as starting point for
new approximative methods, such as WKB \cite{Note2} or linearization around a
given background solution (see the Appendix).

The structure of the Dirac equation suggests that this general strategy can also
be applied to 3+1 dimensions, where both the potential $A_\mu$ and the Dirac
bi-spinor $\psi$ have four components.
Thus, for a given $\psi$, we get four equations for the four components 
$A_\mu$, which can be solved (except in singular cases). 
However, the four constraints $\Im(A_\mu)=0$ assume a form which is far more 
complicated than in 1+1 and 2+1 dimensions.
This renders the identification of real generating functions which correspond 
to the remaining degrees of freedom rather cumbersome. 
The analysis could be simplified by restricting the space-time dependence to 
1+1 and 2+1 dimensions, which should be the subject of further investigations.


\begin{acknowledgments}
R.S.\ acknowledges support by DFG (SFB-TR12) and would like to express special 
thanks to the Perimeter Institute for Theoretical Physics 
and the Mainz Institute for Theoretical Physics (MITP)
for hospitality and support.
\end{acknowledgments}

\appendix 
\section{Perturbed solution}

To find solutions for electric fields that create electron-positron pairs 
(see, e.g., \cite{Schutzhold2008,Monin2010,Dunne2009,Orthaber2011}), 
we use the ansatz
\begin{equation}
  r = \alpha + \beta \sin (m \gamma),
\end{equation}
where the Bogoliubov coefficients $\alpha$ and $\beta$ as well as the 
eikonal function $\gamma$ are slowly varying functions of the light cone 
coordinates.
(We consider 1+1 dimensions for simplicity.)
The main idea here is that $\alpha$ is an exact solution and $\beta$ 
is used to slowly turn on an oscillating perturbation. 
The value of $\beta$ then is related to the pair creation rate.

However, the calculation of $s$ and $q E$ is rather complicated for arbitrary
functions $\alpha$, $\beta$, and $\gamma$ because $s$ depends nonlinearly 
on $r$. 
Hence, as the perturbation should be small, we calculate the electric field 
only up to linear order in $\beta$
\begin{equation}
\label{eq:lin-beta}
E = E^{(\alpha)} + E^{(\beta)} + \mathcal{O}(\beta^2),
\end{equation}
where $q E^{(\alpha)}$ is the unperturbed force of order $\beta^0$ and $q
E^{(\beta)}$ is the first-order perturbation of order $\beta^1$.
Apart from this linearization, we assume that the mass $m$ represents the 
largest energy scale in the problem and thus we employ a large-$m$ expansion 
on top of the approximation in Eq.~\eqref{eq:lin-beta}. 
Expanding $q E^{(\beta)}$ into powers of $m$ and keeping only the highest-order 
term gives
\begin{equation}
\begin{aligned}
q E^{(\beta)} 
&= 
\sqrt{2}\beta\,\frac{\cos (m\gamma)}{s_\alpha \pd{+} \gamma} 
\Bigg[ m^2 (\pd{+} \gamma)^2 (\pd{-} \gamma)^2 
\\
&- \bigg( 
\underbrace{\frac{m}{\sqrt{2}} \frac{\alpha}{s_\alpha}}_{= -q A_+^{(\alpha)}} 
\pd{-} \gamma + 
\underbrace{\frac{m}{\sqrt{2}} \frac{s_\alpha}{\alpha}}_{= -q A_-^{(\alpha)}} 
\pd{+} \gamma \bigg)^2 \Bigg] + \mathcal{O}(m^1),
\end{aligned}
\label{eq:firstorderm}
\end{equation}
with the abbreviation 
\begin{equation}
s_\alpha =\sqrt{c - \int \pd{-} \alpha^2 \intd{x_+}}. 
\end{equation}
($qA_+^{(\alpha)}$ and $qA_-^{(\alpha)}$ are the leading-order
contributions to the vector potential.)
Since $\alpha$, $\beta$, and $\gamma$ are supposed to be slowly varying, 
the leading contribution~\eqref{eq:firstorderm} would be rapidly oscillating 
due to the pre-factor $\cos(m\gamma)$ unless the phase function $\gamma$ 
has a stationary point (see below). 
Of course, such a rapidly oscillating force with a frequency of order $m$ could
well create pairs, but this process would be typically in the perturbative
(multi-photon) regime.
Here, we are interested in non-perturbative phenomena such as the 
Sauter-Schwinger effect and thus we demand that these rapidly oscillating 
contributions are absent -- at least to leading order.
Thus, we require the term of order $m^2$ in $q E^{(\beta)}$ to vanish. 
This is the case if $S = m \gamma$ solves the eikonal equation
\begin{equation}
\frac{m^2}{2} 
= 
\left( \pd{+} S + q A_+^{(\alpha)} \right) \left( \pd{-} S + q A_-^{(\alpha)} \right).
\label{eq:eikonaleq}
\end{equation}
Therefore, this condition can be used to fix $\gamma$ for a given $\alpha$. 
Then, the leading order of $q E^{(\beta)}$ is of order $m^1$
\begin{widetext}
  \begin{equation}
    \begin{aligned}
      q E^{(\beta)} &= \frac{m}{\sqrt{2}} \frac{1}{s_\alpha} \sin (m\gamma) \Bigg\{
      2 (\pd{+} \beta) \bigg[ \frac{\pd{-} \gamma}{\pd{+} \gamma} + \left( \frac{\alpha}{s_\alpha} \frac{\pd{-} \gamma}{\pd{+} \gamma} \right)^2 \bigg]
      +2 (\pd{-} \beta) \bigg[ 1 + \left( \frac{s_\alpha}{\alpha} \right)^2 \frac{\pd{+} \gamma}{\pd{-} \gamma} \bigg] \\
      &+\beta \bigg[
        \frac{\pd{+} \alpha}{\alpha} \bigg( \frac{\pd{-} \gamma}{\pd{+} \gamma} + \left( \frac{\alpha}{s_\alpha} \frac{\pd{-} \gamma}{\pd{+} \gamma} \right)^2 + 2 \left( \frac{s_\alpha}{\alpha} \right)^2 \bigg)
      + \frac{\pd{-} \alpha}{\alpha} \bigg( 2 + \left( \frac{s_\alpha}{\alpha} \right)^2 \frac{\pd{+} \gamma}{\pd{-} \gamma} + \left( \frac{\alpha}{s_\alpha} \right)^2 \left( 2 (\pd{-} \gamma)^2 - \frac{\pd{-} \gamma}{\pd{+} \gamma} \right) \bigg) \\
      &\quad + \pd{+}(\pd{-} \gamma)^2 + \left( \frac{\alpha}{s_\alpha} \right)^2 \frac{\pd{-} \gamma}{\pd{+} \gamma} \pd{+} \left( \frac{\pd{-} \gamma}{\pd{+} \gamma} \right) \bigg] \Bigg\} + \mathcal{O}(m^0),
    \end{aligned}
  \end{equation}
\end{widetext}
where we have used the eikonal equation \eqref{eq:eikonaleq} to simplify 
some expressions. 
If we require this rapidly oscillating term to vanish as well, we get a 
linear first-order partial differential equation for $\beta$. 
However, this linear equation does not have any source term. 
Therefore, a solution where $\beta$ vanishes initially will not
generate any pairs unless the coefficients of $\pd{+} \beta$ and
$\pd{-} \beta$ vanish at some point. 
As those are proportional to
\begin{equation}
\frac{m}{\sqrt{2}} \frac{\alpha}{s_\alpha} \pd{-} \gamma + 
\frac{m}{\sqrt{2}} \frac{s_\alpha}{\alpha} \pd{+} \gamma,
\label{eq:pccond}
\end{equation}
we only obtain pair creation at this level of description if 
\eqref{eq:pccond} vanishes somewhere. 
According to the eikonal equation \eqref{eq:firstorderm}, this in turn 
implies that $\pd{+} \gamma$ or $\pd{-} \gamma$ has to vanish somewhere,
i.e., that the phase function becomes stationary.

\end{document}